\documentclass[preprintnumbers,twocolumn,article,amsmath,amssymb,floatfix,9pt,prd,superscriptaddress,nofootinbib]{revtex4-1}
\usepackage[utf8,latin1]{inputenc}
\usepackage{graphicx}
\usepackage{dcolumn}
\usepackage[dvipsnames]{xcolor}
\usepackage[T1]{fontenc}

\usepackage{mathrsfs}  
\usepackage{cases}
\usepackage{bm}
\usepackage{academicons}
\usepackage{mathtools, nccmath}
\usepackage{fancyhdr}
\usepackage{tikz,xcolor}
\newcommand{\qm}[1]{``#1''}
\usepackage{tensor}
\usepackage[normalem]{ulem}
\usepackage{lipsum}
\usepackage{soul}
\usepackage{cancel}
\usepackage{stackengine,scalerel}
\usepackage{mathabx}
\usepackage{hyperref}
\usepackage{tabularx}
\hypersetup{colorlinks, linkcolor={red},citecolor={blue},urlcolor={blue}}  

\AtBeginDocument{\fontsize{7}{10}\selectfont}

\definecolor{lime}{HTML}{A6CE39}
\DeclareRobustCommand{\orcidicon}{
	\begin{tikzpicture}
	\draw[lime, fill=lime] (0,0) 
	circle [radius=0.16] 
	node[white] {{\fontfamily{qag}\selectfont \tiny ID}};
	\draw[white, fill=white] (-0.0625,0.095) 
	circle [radius=0.007];
	\end{tikzpicture}
	\hspace{-2mm}
}

\fancypagestyle{plain}{%
  \fancyhf{}
  \fancyfoot[C]{\iffloatpage{}{\thepage}}
  }
\foreach \x in {A, ..., Z}{%
	\expandafter\xdef\csname orcid\x\endcsname{\noexpand\href{https://orcid.org/\csname orcidauthor\x\endcsname}{\noexpand\orcidicon}}
}

\begin{document}

\title[Coherent Quantum-State Corrections in Reissner-Nordstr\"{o}m Black Holes: Photon Dynamics, Shadow Casts, and Energy Backreaction]{Coherent quantum state corrections in RN--black holes: Photon dynamics, shadow casts, and energy backreaction}

\author{A. Errehymy\orcidA{}}%
\email{abdelghani.errehymy@gmail.com (Corresponding author)}
\affiliation{Astrophysics Research Centre, School of Mathematics, Statistics and Computer Science, University of KwaZulu-Natal, Private Bag X54001, Durban 4000, South Africa}
\affiliation{Center for Theoretical Physics, Khazar University, 41 Mehseti Str., Baku, AZ1096, Azerbaijan}
\affiliation{Jadara Research Center, Jadara University, Irbid 21110, Jordan}

\author{Y. Khedif\orcidB{}}%
\email{youssef.khedif@gmail.com}
\affiliation{Department of Physics, Faculty of Sciences A\"{i}n Chock, Laboratory of Mechanics and High Energy Physics, University Hassan II, P.O. Box 5366, 20100, Maarif Casablanca, Morocco}

\author{M. Daoud\orcidC{}}%
\email{m$_{}$daoud@hotmail.com}
\affiliation{Department of Physics, Faculty of Sciences, Ibn Tofail University, P.O. Box 133, Kenitra 14000, Morocco}
\affiliation{Abdus Salam International Centre for Theoretical Physics, Miramare, Trieste 34151, Italy}

\author{B. Turimov\orcidD{}}%
\email[]{bturimov@astrin.uz}
\affiliation{Engineering school, Central Asian University, Milliy bog Str.264, Tashkent, 111221,Uzbekistan}
\affiliation{University of Tashkent for Applied Sciences, Gavhar Str.1, Tashkent, 100149, Uzbekistan}
\affiliation{Ulugh Beg Astronomical Institute, Astronomy St. 33, Tashkent 100052, Uzbekistan}   

\author{S. Usanov\orcidE{}}%
\email[]{sm.usanov@kiut.uz}
\affiliation{Kimyo International University in Tashkent, Shota Rustaveli Str. 156, Tashkent 100121, Uzbekistan}

\author{Z. Yasakov\orcidF{}}%
\email[]{zikrillo87@mail.ru}
\affiliation{Samarkand State University of Architecture and Construction, Lolazor Street 70, 140147, Samarkand, Uzbekistan}

\author{F. Turaev\orcidI{}}%
\email[]{farhodjon9618@mail.ru}
\affiliation{Alfraganus University, Yukori Karakamish Str. 2a, Tashkent 100190, Uzbekistan}

\date{\today} 

\begin{abstract}
{\footnotesize { We revisit the Reissner-Nordstr\"{o}m (RN) geometry within the coherent quantum black hole framework, introducing a version that remains well-behaved at short distances while preserving the expected classical behavior at large scales. The key ingredient is a smooth Gaussian regularization, which replaces the usual sharp cutoff and {suppresses} the unphysical oscillations. This construction introduces a characteristic core radius $R_{\rm s}$, constrained relative to the mass $M$ and charge $Q$ to ensure either a {Ricci-regular} center or a softened integrable singularity. Near the origin, the spacetime exhibits distinct behaviors depending on the choice of parameters. When $R_{\rm s}$ satisfies the critical relation $R_{\rm s}^* = \sqrt{\pi}\,R_{\rm Q}^2 / R_{\rm M}$, the geometry is {Ricci-regulFar}, while deviations from this value produce a milder, integrable singularity. {We clarify that for generic parameters, the Kretschmann scalar retains a $1/r$ divergence even at $R_{\rm s}^*$, so the center is not fully regular in the sense of finite curvature invariants.} The effective stress-energy tensor can be interpreted as an anisotropic fluid, whose density and pressure smoothly interpolate between the quantum-dominated core and the classical RN region at large radii. The Misner-Sharp mass accumulates gradually from the center, recovering the ADM mass asymptotically. The quantum scale also modifies the horizon structure, giving rise to standard, extremal, or horizonless configurations depending on the ratios $\alpha=R_{\rm Q}/R_{\rm M}$ and $\beta=R_{\rm s}/R_{\rm M}$. Photon motion is correspondingly affected: the photon sphere moves inward, light deflection and Shapiro delay are slightly reduced, and the resulting shadow appears more compact and smoother, reflecting the quantum-corrected geometry. These features illustrate how parameter choices control both regularity and observable signatures. Altogether, this work revisits charged black holes with quantum modifications, free from spurious oscillations and offering measurable consequences. }}
\end{abstract}

\maketitle

\section{Introduction:} \label{Sec:1}
Among the most striking outcomes of Einstein's theory of gravitation are black holes, whose existence is now firmly supported by multiple observational channels. The detection of gravitational waves from merging compact binaries by the LIGO-Virgo-KAGRA collaboration \cite{LIGOScientific2016,LIGOScientific2025}, the imaging of the shadow structures of M87$^*$ and Sgr A$^*$ by the Event Horizon Telescope \cite{EventHorizonTelescope2019,EventHorizonTelescope2022}, and the electromagnetic emission from accreting matter \cite{Nampalliwar2018} together provide a coherent picture of these extreme systems. Recent astrophysical data-analysis techniques have further expanded this observational landscape, including gamma-ray blazar identification and high-accuracy Doppler-shift measurements \cite{Xiang:2026, Zhang:2026}. From a theoretical standpoint, black holes also expose one of the deepest features of general relativity: the appearance of spacetime singularities. This is already evident in the Schwarzschild solution, obtained in 1916 \cite{Schwarzschild1916}, which describes the gravitational field outside a static, spherically symmetric body. The corresponding geometry contains a curvature divergence at $r=0$, where invariants such as the Kretschmann scalar blow up, together with an event horizon at $r=2M$ that defines a \qm{no-escape region}, acting as a one-way causal boundary.

Although such behavior was initially associated with the idealized symmetry of the model, later developments showed that it is not merely an artifact of spherical symmetry. The singularity theorems of Penrose and Hawking \cite{Hawking1973-book} established that these features arise generically in gravitational collapse. In particular, Penrose's 1965 theorem \cite{Penrose1965} showed---without relying on symmetry---that the formation of a closed trapped surface leads to geodesic incompleteness \cite{Landsman2021,Landsman2022}. In this sense, the theorem concerns the inability to extend causal trajectories indefinitely rather than directly proving the existence of an observable black hole. The usual interpretation in terms of black-hole formation is then connected with the cosmic censorship conjecture \cite{Penrose1969}, according to which singularities produced by gravitational collapse remain hidden behind event horizons and are therefore inaccessible to distant observers \cite{Wald-book1984}.

Seen from this perspective, singularities mark a genuine limitation of classical general relativity. At such points, the theory ceases to provide a regular description of spacetime, while physical quantities can diverge and predictive power is lost \cite{Hawking1979-path}. This has motivated a broad search for alternatives that retain the successful predictions of general relativity while modifying its behavior in extreme regimes. Observational studies of compact objects have likewise kept open the possibility of departures from the standard black-hole picture \cite{Berti2015,Barack2018,LISA2022}. Modified theories of gravity provide one possible route in this direction \cite{Capozziello:2011et}, while scalar--fermion interactions have also been explored in alternative gravitational frameworks \cite{Razina:2010bj,Razina:2011wv}. Related mathematical structures have further connected gravitational models with nonlinear differential equations \cite{Esmakhanova:2011ar}.

{Several possibilities have emerged within this broader program. Horizonless exotic compact objects have been proposed as black-hole mimickers in strong-gravity environments \cite{Cardoso-Pani2019}. Regular black-hole models, in contrast, retain horizons while replacing the central singularity with a finite or otherwise controlled core. Such constructions have appeared in higher-dimensional settings, including Myers--Perry geometries under appropriate conditions \cite{Emparan2008,Myers2011}, as well as in a variety of four-dimensional regular spacetimes \cite{Ansoldi2008,Lan2023,Carballo-Rubio2025}. More recent studies have examined nonsingular and quantum-corrected black holes through orbital motion, photon trajectories, lensing, and shadow observables \cite{Bambi:2023try,Yang:2026kmk,Errehymy:2025zuj}. Other analyses have considered scalar-hairy black holes and black holes surrounded by dark matter or quintessence \cite{Liang:2026vls,Errehymy:2025pfr,Errehymy:2025ada}.}

{The strong-field motion of matter provides another useful way of probing these geometries. Charged and magnetized particles have been studied around black holes in conformal gravity, Einstein--\AE ther gravity, and Einstein--Maxwell-scalar theory \cite{Narzilloev:2020jqn,Rayimbaev:2020gqj,Zahid:2022zzt}. Their orbital dynamics can also be connected with quasi-periodic oscillations observed in accreting compact objects \cite{Rayimbaev:2022pzr,Rayimbaev:2024abz,Rayimbaev:2024ywi}. Similar investigations have been carried out for Simpson--Visser and Black--bounce--RN spacetimes, including braneworld scenarios \cite{Vrba:2023byq,Mitra:2024olo}. Magnetic fields, particle spin, and particle collisions introduce further effects into the motion near charged and rotating black holes \cite{Jumaniyozov:2024hlg,Abdulxamidov:2023jfq,Narzilloev:2021bst}. These studies illustrate how particle dynamics and QPOs can complement photon-based observables when testing strong gravitational fields.}

Charged black holes offer a particularly useful arena in which to examine these questions more closely. The RN solution, obtained from the Einstein--Maxwell equations, describes a static, spherically symmetric black hole carrying electric charge and is uniquely selected---under appropriate assumptions---by a no-hair theorem \cite{Israel:1967za}. Although realistic astrophysical black holes are expected to be nearly neutral, the RN geometry remains a valuable theoretical laboratory because of its rich causal structure \cite{Hawking:1973uf}. In addition to the curvature singularity at $r=0$, it possesses an inner (Cauchy) horizon whose stability is compromised by perturbations \cite{PhysRevLett.63.1663,PhysRevD.41.1796}, with the resulting growth of the internal mass commonly associated with the mass-inflation mechanism.

The persistence of these singular structures has long motivated the search for a quantum description of gravity. Numerous regular black-hole models have therefore been proposed, with the aim of replacing the singular core by an effective geometry in which curvature remains finite \cite{Bambi:2023try}. However, some of these constructions still contain inner Cauchy horizons, leaving important questions concerning their stability. An alternative is provided by the coherent-state approach to quantum black holes \cite{Casadio:2016zpl,Casadio:2017cdv,Casadio:2021eio}. Rather than introducing an ad hoc modification of the classical metric, this framework starts from quantum-mechanical states and is closely related to the quantum $N$-portrait proposed by Dvali and Gomez \cite{Dvali:2011aa}. In this picture, a black hole is viewed as a macroscopic quantum system composed of a large number of weakly interacting gravitons and matter quanta, while the coherent state provides an effective description of their collective behavior.

In this paper, we revisit the RN black hole within the coherent quantum-state framework, focusing on how a smooth Gaussian regularization changes the geometry near the core. We study the conditions under which the center becomes {Ricci-regular} or develops a mild, integrable singularity, derive the corresponding anisotropic stress-energy tensor, and examine the resulting Misner-Sharp mass profile. We then investigate how the quantum scale affects the horizon structure, photon trajectories, and black-hole shadow. Particular attention is given to the roles of the mass, charge, and core radius in controlling the regularity of the geometry and its observable properties.

The paper is structured as follows. In Sect. \ref{sec-II}, we revisit the RN geometry within the coherent quantum framework. This includes: (\ref{sSec:A}) constructing the effective potential from coherent quantum states, (\ref{sSec:B}) implementing a smooth Gaussian regularization and examining the resulting geometry, (\ref{sSec:C}) analyzing the behavior at the origin, (\ref{sSec:D}) deriving the effective anisotropic source and its stress-energy tensor, and (\ref{sSec:E}) exploring near-core properties and the Misner-Sharp mass. Sect. \ref{sec-III} focuses on photon motion around these quantum-corrected black holes, covering (\ref{ssSec:A}) inner and outer horizons, (\ref{ssSec:B}) light propagation in the modified spacetime, and (\ref{ssSec:C}) Shapiro time delay. In Sect. \ref{sec-IV}, we study the shadow cast by the black holes, while Sect. \ref{sec-V} discusses energy accumulation and backreaction effects. Finally, Sect. \ref{sec-VI} presents a discussion of our findings and the concluding remarks.

\section{Refined quantum description of the RN geometry}
\label{sec-II}

We start by introducing coherent quantum states and the resulting effective potential (Subs.~\ref{sSec:A}). Then, we explore the geometry obtained through Gaussian smoothing (Subs.~\ref{sSec:B}). Next, we analyze the singularity or regularity at the origin (Subs.~\ref{sSec:C}). After that, we present the effective anisotropic source and its stress--energy tensor (Subs.~\ref{sSec:D}). Finally, we discuss the near-core behavior and the Misner--Sharp mass (Subs.~\ref{sSec:E}).

\subsection{From coherent quantum states to the effective potential}\label{sSec:A}

The purpose of this section is to revisit the quantum-corrected RN spacetime and eliminate the unphysical oscillatory behavior reported in~\cite{Casadio:2022ndh}. This is achieved by implementing the smooth regularization scheme introduced in~\cite{Feng:2024nvv}. Rather than retracing the full derivation, we concentrate on the modifications that lead to a more consistent result. Within the {\em coherent quantum black hole} picture~\cite{Casadio:2016zpl,Casadio:2017cdv,Casadio:2021eio}, a classical spacetime emerges as an averaged description of an underlying quantum state. For a static, spherically symmetric and asymptotically flat configuration, the line element can be written as
\begin{equation}\label{eq:fmetric}
ds^2 = -h(r)\,dt^2 + \frac{dr^2}{h(r)} + r^2 d\Omega^2 \, ,
\end{equation}
where $r$ denotes the areal radius and the horizon structure is determined by the condition $h(r)=0$ or equivalently $h(r)= 1+2V(r)$. In this framework, the gravitational potential $V(r)$ is not fundamental but arises as the expectation value of a quantum operator,
\(V(r) = \langle g | \widehat{\Phi}(r) | g \rangle \), with $\widehat{\Phi}$ representing an effective scalar degree of freedom encoding the collective quantum dynamics.

For the classical RN spacetime, the gravitational potential is given by
\begin{equation}
V_{\text{RN}}(r)
=
-\frac{G_{\rm N} M}{r}
+\frac{G_{\rm N}Q^2}{2r^2},
\end{equation}
where \(M\) and \(Q\) correspond to the ADM mass and electric charge of the compact object.

\begin{figure}
\centering
\includegraphics[width=5.9cm,height=4.9cm]{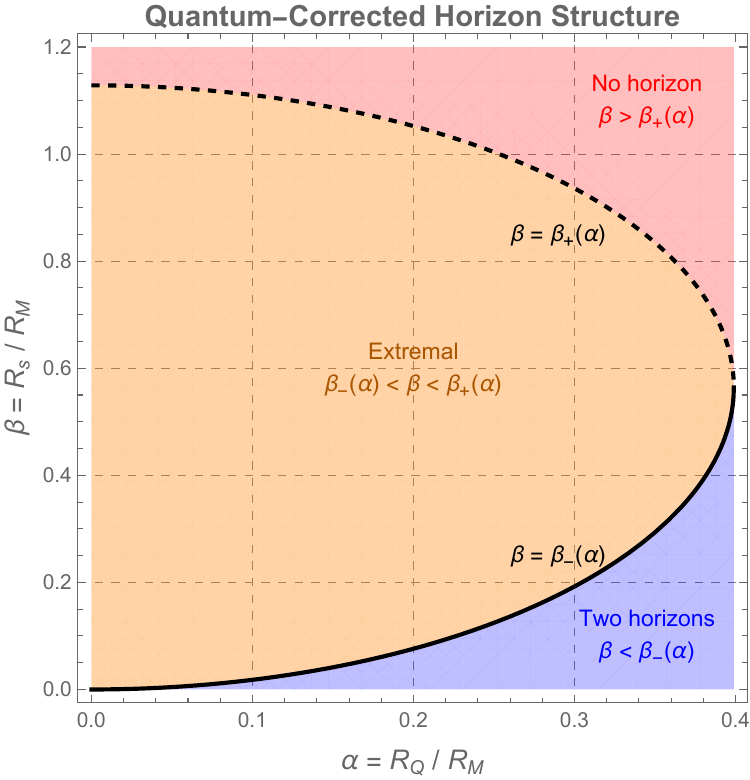}%
    \caption{\scriptsize This diagram illustrates shows how quantum effects change the horizon structure in the $\alpha$--$\beta$ plane, where $\alpha = R_Q / R_M$ and $\beta = R_s / R_M$ {(both dimensionless)}. In the blue region, spacetimes have two separate horizons. The orange area marks extremal cases where the inner and outer horizons merge. In the red region, no horizons exist and the singularity is exposed. Solid and dashed black curves indicate the critical boundaries $\beta_{\pm}(\alpha)$ from Eq.~(\ref{eq31}), which separate these different behaviors. Labels inside each region clarify the conditions for the horizons, providing a clear picture of how quantum corrections shift and smooth-the classical RN structure.
    } \label{fig:horizons_RN}
\end{figure}

\subsection{Smooth Gaussian regularization and resulting geometry}\label{sSec:B}

As discussed in~\cite{Casadio:2022ndh}, the exact RN potential cannot be reproduced at the quantum level without encountering ultraviolet divergences generated by high-momentum modes. A natural way to regularize the theory is therefore to introduce a momentum cutoff at the scale \(k\sim R_{\rm s}^{-1}\), with \(R_{\rm s}\) interpreted as the characteristic size of the quantum core. The regularized potential can then be expressed as
\begin{equation}
V_{RN}(r)
=
\int_{|\vec{k}|<R_s^{-1}}
\frac{d^3k}{(2\pi)^3}\,
\widetilde{V}_{RN}(k)\,
e^{i\vec{k}\cdot\vec{r}} .
\end{equation}

For a spherically symmetric configuration, one may write \(d^3k=k^2dk\,d\Omega\), so that the above relation becomes \cite{Antonelli:2025mcv}
\begin{equation}
V_{RN}(r)
=
\frac{1}{2\pi^2}
\int_0^{R_s^{-1}}
k^2\,dk\,
\widetilde{V}_{RN}(k)\,
j_0(kr),
\end{equation}
where \(j_0(kr)=\frac{\sin(kr)}{kr}\) is the spherical Bessel function of order zero, obtained from the angular integral
\begin{equation}
\int d\Omega\,e^{i\vec{k}\cdot\vec{r}}
=
4\pi\,j_0(kr).
\end{equation}
\begin{figure}
\begin{center}
\includegraphics[width=7.9cm,height=4.9cm]{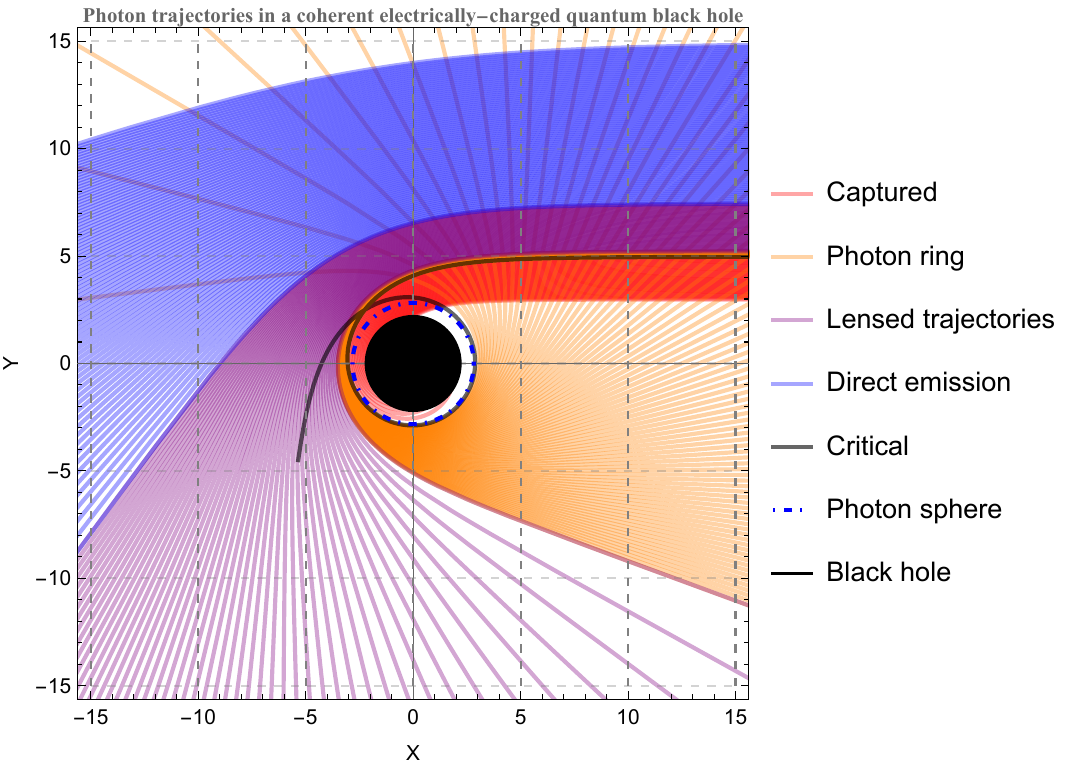}
\end{center}
\caption{\scriptsize  The plot shows photon trajectories around a coherent electrically-charged quantum black hole with \(\alpha=0.45\) and \(\beta=0.25\), corresponding to \(R_{\rm M}=2\), \(R_{\rm Q}=0.5\), and \(R_{\rm s}=0.9\). Red curves depict photons that are captured by the black hole, while orange curves trace paths that loop near the photon sphere (blue dot-dashed circle), forming the photon ring. Purple curves show strongly bent trajectories that circle the black hole before escaping, and blue curves represent nearly straight, weakly deflected paths. The black curve highlights the critical trajectory that separates captured and escaping photons, and the central black disk marks the black hole region.
}\label{fig:trajectory}
\end{figure}
Although the sharp cutoff removes the ultraviolet divergence, it also produces oscillatory terms involving sine-integral and cosine functions. These oscillations are not associated with any physical property of the spacetime; instead, they arise from the abrupt truncation of the momentum modes. To avoid this issue, it is more convenient to replace the sharp cutoff with a smooth suppression of large momenta. Following~\cite{Feng:2024nvv}, we introduce a Gaussian damping factor and define the quantum-corrected potential as
\begin{equation}
V_{\text{RN}}^{\text{q}}(r)
=
\int_{0}^{\infty}
\frac{d^3k}{(2\pi)^3}\,
\tilde{V}_{\text{RN}}(k)\,
e^{-k^2R_{\rm s}^2/4}\,
e^{i\vec{k}\cdot\vec{r}}.
\end{equation}
{The factor \(1/4\) in the exponent \(e^{-k^2R_{\rm s}^2/4}\) is a standard convention chosen so that the Gaussian width in momentum space matches the characteristic scale \(R_{\rm s}\) in position space. Specifically, the Fourier transform of a Gaussian with width \(\sigma = R_{\rm s}/\sqrt{2}\) yields the damping factor \(e^{-k^2R_{\rm s}^2/4}\), ensuring that the position-space smoothing kernel has a simple form and that the real-space cutoff scale is precisely \(R_{\rm s}\).}

After carrying out the angular integration, the potential can be written in the compact form
\begin{equation}\label{PotentV}
V_{\text{RN}}^{\text{q}}(r)
=
-\frac{G_{\rm N}M}{r}\,
\operatorname{erf}\!\left(\frac{r}{R_{\rm s}}\right)
+\frac{G_{\rm N}Q^2}{2r^2}\,
\mathcal{D}\!\left(\frac{r}{R_{\rm s}}\right),
\end{equation}
where \(\operatorname{erf}(x) = \frac{2}{\sqrt{\pi}} \int_0^x e^{-t^2}dt\) denoting the error function and \(\mathcal{D}(x)
=
\frac{2x}{\sqrt{\pi}}\,e^{-x^2}
\int_0^x e^{t^2}dt\) representing a Dawson-type function.

The associated lapse function becomes
\begin{equation}
h(r)
= 1 - \frac{2G_{\rm N} M}{r}\,\operatorname{erf}\left(\frac{r}{R_{\rm s}}\right)
+ \frac{G_{\rm N} Q^2}{r^2}\,\mathcal{D}\left(\frac{r}{R_{\rm s}}\right) \, .
  \end{equation}
 {Here we define the characteristic scales $R_{\rm M}=2G_{\rm N}M$ and $R_{\rm Q}=\sqrt{G_{\rm N}}\,Q$,} this can be rewritten as
  \begin{equation} \label{eq:quantumf}
  h(r)
  = 1 - \frac{R_{\rm M}}{r}\,\operatorname{erf}\left(\frac{r}{R_{\rm s}}\right)
+ \frac{R_{\rm Q}^2}{r^2}\,\mathcal{D}\left(\frac{r}{R_{\rm s}}\right) \, .
  \end{equation}

This construction leads to a smooth geometry free from the artificial oscillations encountered previously. Quantum corrections are significant only in the vicinity of the core scale $r \sim R_{\rm s}$, while at large distances the classical RN solution is rapidly recovered due to the asymptotic behavior of the regulating functions. In the limiting case $R_{\rm s} \to 0^{+}$, the damping disappears and the standard RN spacetime is fully restored, including its central singularity.

\begin{figure}
\begin{center}
\includegraphics[width=7.9cm,height=4.9cm]{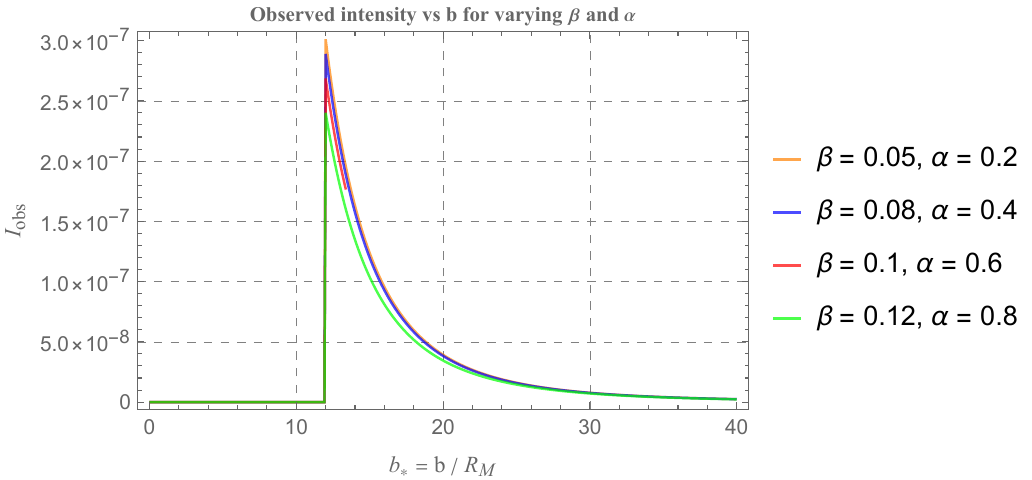}
\end{center}
\caption{\scriptsize Variation of the observed intensity \(I_{\rm obs}\) with the normalized impact parameter \(b_* = b / R_M\) for {four} combinations of smearing \(\beta\) and charge \(\alpha\). {The curves correspond to \((\beta, \alpha) = (0.05, 0.2)\) (blue solid), \((0.08, 0.4)\) (orange dashed), \((0.10, 0.6)\) (green dot-dashed), and \((0.12, 0.8)\) (red dotted).} Each line illustrates how the interplay of the object's metric and emission profile shapes the intensity seen by a distant observer. Distinct colors and line patterns are used to separate the {four} \((\beta, \alpha)\) scenarios, showing how stronger charge or larger smearing alters the brightness distribution around the source.
}\label{fig:intensity}
\end{figure}

\subsection{Behavior at the Origin}\label{sSec:C}

The quantum-corrected RN metric \(h(r)=1+2V_{\rm RN}^{\rm q}(r) \), being analytic, admits a Taylor expansion around $r=0$:
\begin{equation}
h(r) = \sum_{k=0}^{\infty} h_k\, r^k, \quad \text{where each coefficient is } h_k = \frac{1}{k!} \left. \frac{d^k h}{dr^k} \right|_{r=0}.
\end{equation}

In terms of these coefficients, the curvature invariants read near the origin:
\begin{align}
R &= -\frac{2(h_0-1)}{r^2}-\frac{6 h_1}{r}-12 h_2 + \mathcal{O}(r), \\
\tensor{R}{^\mu_\nu}\tensor{R}{^\nu_\mu} &= \frac{2(h_0-1)^2}{r^4} + \frac{8 h_1(h_0-1)}{r^3} + \frac{2[6 h_2(h_0-1)+5 h_1^2]}{r^2} \notag\\
&\quad + \frac{4[4 h_3(h_0-1)+9 h_1 h_2]}{r} \notag\\
&\quad + 4[5 h_4(h_0-1)+14 h_1 h_3+9 h_2^2] + \mathcal{O}(r), \\
\mathcal{K} &= \frac{4(h_0-1)^2}{r^4} + \frac{8 h_1(h_0-1)}{r^3} + \frac{8[h_2(h_0-1)+h_1^2]}{r^2} \notag\\
&\quad + \frac{8[h_3(h_0-1)+3 h_1 h_2]}{r} \notag\\
&\quad+ 8[h_4(h_0-1)+4 h_1 h_3+3 h_2^2] + \mathcal{O}(r),
\end{align}
where \(R\) is the Ricci scalar, \(\tensor{R}{^\mu_\nu}\tensor{R}{^\nu_\mu}\) the squared Ricci tensor, and \(\mathcal{K}\) the Kretschmann scalar. These invariants remain finite if \(h_0=1\) and \(h_1=0\). 

{For the quantum-corrected RN metric~\eqref{eq:quantumf}, the coefficients are}
\begin{eqnarray*}
h_0 &=& 1 + \frac{2}{R_{\rm s}} \Big({\frac{R_{\rm Q}^2}{R_{\rm s}}} - \frac{R_{\rm M}}{\sqrt{\pi}}\Big), \nonumber\\
h_1  &=& 0, \nonumber\\
h_2  &=& -\frac{4}{3 R_{\rm s}^3} \Big(\frac{R_{\rm Q}^2}{R_{\rm s}} - \frac{R_{\rm M}}{2\sqrt{\pi}}\Big), \nonumber\\
h_3  &=& \frac{2}{R_{\rm s}^4} \Big(\frac{R_{\rm Q}^2}{R_{\rm s}} - \frac{R_{\rm M}}{2\sqrt{\pi}}\Big),\notag\\
h_4 &=& -\frac{2}{5 R_{\rm s}^5} \Big(\frac{R_{\rm Q}^2}{R_{\rm s}} - \frac{R_{\rm M}}{2\sqrt{\pi}}\Big).
\end{eqnarray*}

{Near \(r=0\), the curvature invariants can be expressed as}
\begin{align}
R &\sim -\frac{2(h_0-1)}{r^2} - 12 h_2 + \dots, \\
\tensor{R}{^\mu_\nu}\tensor{R}{^\nu_\mu} &\sim \frac{2(h_0-1)^2}{r^4} + \frac{8 h_3(h_0-1)}{r} + \dots, \\
\mathcal{K} &\sim \frac{4(h_0-1)^2}{r^4} + \frac{8 h_3(h_0-1)}{r} + \dots,
\end{align}

\begin{figure}
\begin{center}
\includegraphics[width=5.9cm,height=5.9cm]{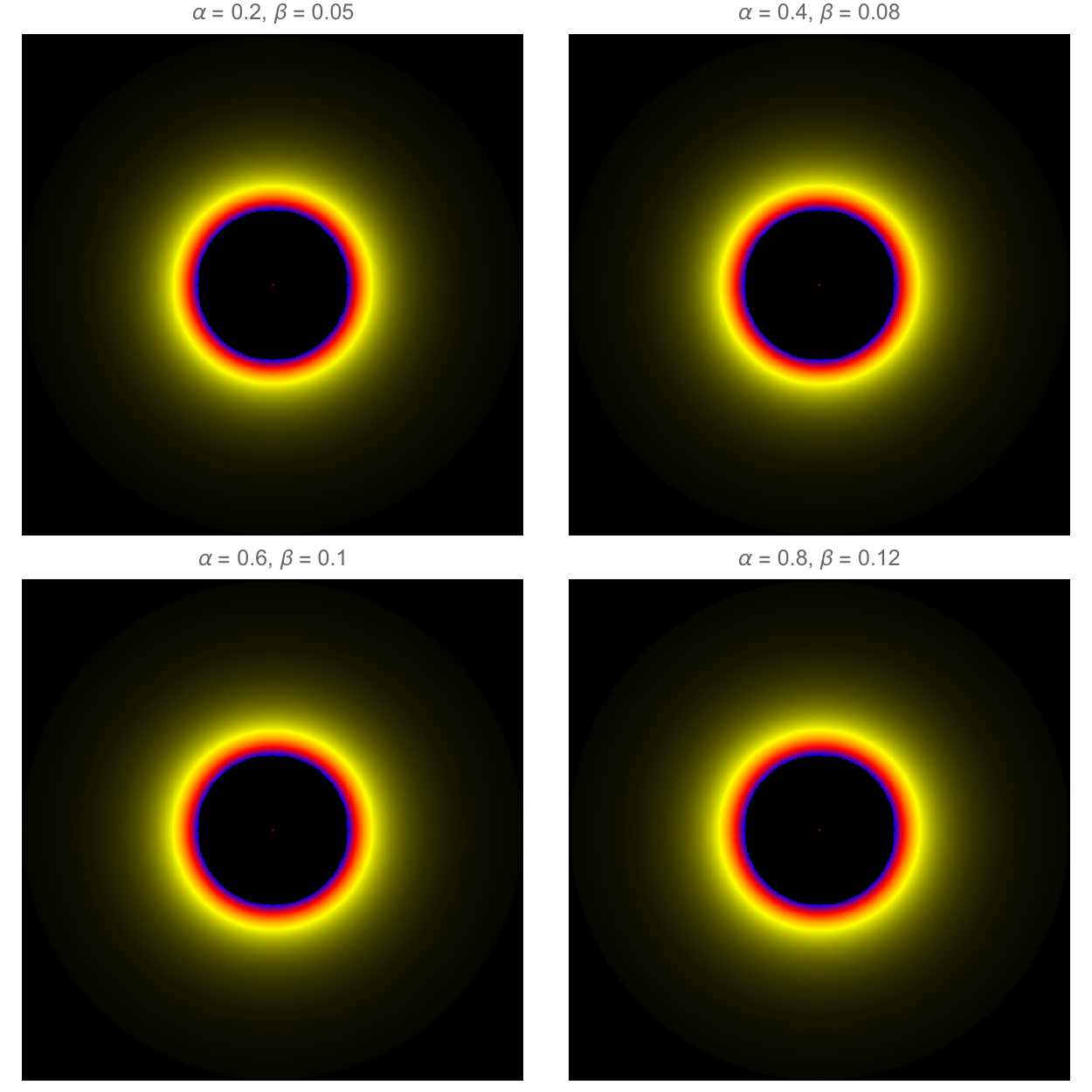}
\end{center}
\caption{\scriptsize  Two-dimensional maps of the observed intensity for coherent electrically-charged quantum black holes, shown for {four} combinations of charge and smearing parameter: {\((\alpha, \beta) = (0.2, 0.05), (0.4, 0.08), (0.6, 0.10), (0.8, 0.12)\)}. The emission is computed in physical units with \(R_M = 2\), an inner cutoff at \(x_{\rm in} = 6\), and an outer boundary at \(x_{\rm out} = 20\). Each panel displays the intensity distribution in the observer's image plane, where the central dark region corresponds to the shadow and the surrounding bright structure traces photon trajectories. As \(\alpha\) increases, the emission becomes more extended and the bright ring shifts outward, while larger \(\beta\) smooths the intensity, reducing sharp contrasts. The color scale (black-yellow-red-blue) highlights the transition from low to high intensity across the image.
}\label{fig:shadow}
\end{figure}

{We now distinguish two situations:}

\begin{itemize}
\item \textbf{{Ricci-regular center:}} {Imposing \(h_0 = 1\) gives}
\begin{equation}\label{eq:regular_condition}
{R_{\rm s} = R_{\rm s}^* \equiv \frac{\sqrt{\pi} R_{\rm Q}^2}{R_{\rm M}}.}
\end{equation}
{At this critical value, the leading \(1/r^4\) terms in the curvature invariants vanish. However, the subleading \(1/r\) terms, proportional to \(h_3(h_0-1)\), remain nonzero for generic parameters. Therefore, while the Ricci scalar becomes finite, the Kretschmann scalar still diverges as \(1/r\) unless an additional fine-tuning condition \(h_3 = 0\) is imposed. This would require}
\begin{equation}
{\frac{R_{\rm Q}^2}{R_{\rm s}} - \frac{R_{\rm M}}{2\sqrt{\pi}} = 0 \quad \Longrightarrow \quad R_{\rm s} = \frac{2\sqrt{\pi} R_{\rm Q}^2}{R_{\rm M}},}
\end{equation}
{which is incompatible with \(R_{\rm s} = R_{\rm s}^*\) unless \(R_{\rm M} = 0\). Thus, for physically relevant parameters, the center is \emph{Ricci-regular} but not fully regular in the sense of finite Kretschmann scalar.}

\item \textbf{{Integrable singularity:}} {If \(h_0 \neq 1\) (\(R_{\rm s} \neq R_{\rm s}^*\)), the origin hosts an integrable singularity~\cite{Lukash:2013ts}. Curvature invariants diverge more mildly than in the classical RN case, while integrals of energy density and pressure remain finite. We emphasize that ``integrable'' here refers to both the curvature invariants and the energy density: their integrals over finite volumes remain well-defined, even though the pointwise values diverge.}
\end{itemize}

{In the limit \(R_{\rm s} \to 0^+\), the classical RN spacetime is recovered, including its central singularity. We note that the expansion coefficients \(h_k\) carry implicit \(R_{\rm s}\)-dependence, so the limit must be taken after evaluating the coefficients at the physical value of \(R_{\rm s}\).}

{These findings align with those in~\cite{Casadio:2022ndh}, showing that the ultraviolet regularization does not alter the singularity structure at the origin.}

\begin{figure}
\begin{center}
\includegraphics[width=5.9cm,height=5.9cm]{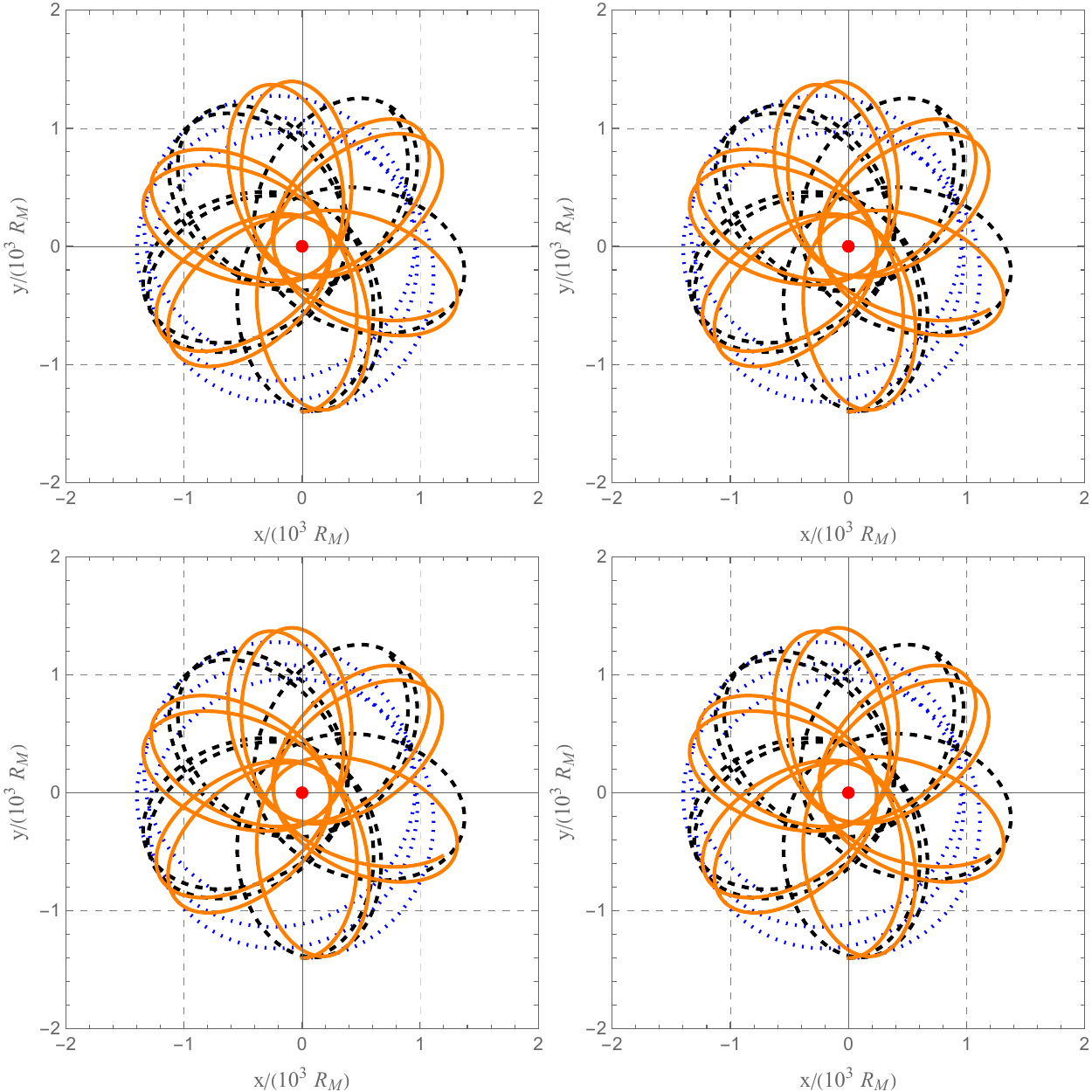}
\end{center}
\caption{\scriptsize  The plot presents the motion of a test particle around a quantum-corrected RN black hole for three different values of \(\alpha = 0.2, 0.5, 0.8\). Each panel shows the particle's trajectory in the \(x\)--\(y\) plane, starting from the initial position \((0, -1.4)\) with velocity directed along the positive \(x\)-axis. As the particle moves under the combined gravitational and electrostatic quantum-corrected potential, its path curves and bends differently depending on the value of \(\alpha\). The three trajectories are color-coded---blue dotted for \(\alpha = 0.2\), black dashed for \(\alpha = 0.5\), and orange solid for \(\alpha = 0.8\)---making it easy to see how increasing \(\alpha\) intensifies the effect of the quantum corrections. The black hole is marked by a red point at the origin, and the axes are scaled by \(10^3 R_M\). This arrangement highlights the gradual deformation of particle paths as the strength of the quantum-corrected potential increases, providing an intuitive visual sense of how quantum effects modify classical trajectories near the black hole.
}\label{fig-orbit}
\end{figure}

\subsection{Effective anisotropic source and stress--energy tensor}\label{sSec:D}

The quantum-corrected metric~\eqref{eq:fmetric} with~\eqref{eq:quantumf} is not a vacuum solution, much like the standard RN case. Instead, it can be interpreted as being sourced by an effective anisotropic fluid, whose profile is shaped by the regularization of the geometry at small $r$.

The stress--energy tensor can be written as
\begin{equation}
\tensor{T}{^\mu_\nu}=\frac{\tensor{G}{^\mu_\nu}}{8\pi G_{\rm N}}
=\text{diag}\left(-\varrho^{\rm q},\,p_r^{\rm q},\,p_t^{\rm q},\,p_t^{\rm q}\right),
\end{equation}
where the components follow from the metric function as

\begin{align}
\varrho^{\rm q}(r)=-\mathcal{P}_r^{\rm q}(r)
&=\varrho_{\rm M}\,A\left(\frac{r}{R_{\rm s}}\right)
-2\,\varrho_{\rm Q}(r)\,B_r\left(\frac{r}{R_{\rm s}}\right),\\
\mathcal{P}_\theta^{\rm q}(r)=\mathcal{P}_\phi^{\rm q}(r)
&=\varrho_{\rm M}\,A\left(\frac{r}{R_{\rm s}}\right)
-2\,\varrho_{\rm Q}(r)\,B_t\left(\frac{r}{R_{\rm s}}\right),
\end{align}
with 
\begin{eqnarray*}
A\left(\frac{r}{R_{\rm s}}\right)&=&e^{-\frac{r^2}{R_{\rm s}^2}},\nonumber\\
B_r\left(\frac{r}{R_{\rm s}}\right)&=&\frac{r^2}{R_{\rm s}^2}\left(1-2\frac{r}{R_{\rm s}}\mathcal{D}\left(\frac{r}{R_{\rm s}}\right)\right),\nonumber\\
B_t\left(\frac{r}{R_{\rm s}}\right)&=&\frac{r^4}{R_{\rm s}^4}\left[\frac{R_{\rm s}}{r}-\left(2-\frac{R_{\rm s}^2}{r^2}\right)\mathcal{D}\left(\frac{r}{R_{\rm s}}\right)\right],\nonumber\\ \varrho_{\rm M}&=&\frac{M}{2\pi^{3/2}R_{\rm s}^3}, 
\quad 
\varrho_{\rm Q}(r)=\frac{Q^2}{8\pi r^4}.
\end{eqnarray*}

\subsection{Near-core behavior and Misner--Sharp mass}\label{sSec:E}

At large radii ($r\gg R_{\rm s}$), the usual electromagnetic contribution is recovered. In contrast, close to the origin one finds
\begin{eqnarray}
\varrho^{\rm q}(r)=-\mathcal{P}_r^{\rm q}(r)&\sim&
\left(\varrho_{\rm M}-2\varrho_{\rm Q}(R_{\rm s})\right)\frac{R_{\rm s}^2}{r^2},\nonumber\\
\mathcal{P}_\theta^{\rm q}(r)=\mathcal{P}_\phi^{\rm q}(r)&\sim& \varrho_{\rm M}-4\varrho_{\rm Q}(R_{\rm s}),
\end{eqnarray}
indicating that the density and radial pressure typically diverge as $r\to0$, while the tangential pressure remains finite. Nevertheless, these quantities are still locally integrable, so their total contributions over finite volumes remain well-defined. This feature will play a role in assessing the nature of the central singularity.

{Finally, the Misner--Sharp mass~\cite{Misner:1964je,Faraoni:2015ula} reads}
\begin{equation}\label{eq:mr}
m^{\rm q}(r)
=4\pi\int_0^r \varrho^{\rm q}(x)x^2dx
= M\,\operatorname{erf}\left(\frac{r}{R_{\rm s}}\right)
-\frac{Q^2}{R_{\rm s}}\mathcal{D}\left(\frac{r}{R_{\rm s}}\right),
\end{equation}
{For any fixed \(r > 0\), the pointwise limit as \(R_{\rm s} \to 0^+\) is}
\begin{equation}
{\lim_{R_{\rm s}\to0^+} m^{\rm q}(r) = M,}
\end{equation}
{since \(\operatorname{erf}(r/R_{\rm s}) \to 1\) and \(\mathcal{D}(r/R_{\rm s}) \sim R_{\rm s}/(2r) \to 0\). The classical RN expression \(M - Q^2/(2r)\) is recovered only in the distributional sense, where the charge contribution becomes localized at the origin as a delta-function singularity. This distinction between pointwise and distributional limits is important for understanding the nature of the classical limit.}

\begin{figure}
\begin{center}
\includegraphics[width=7.9cm,height=4.9cm]{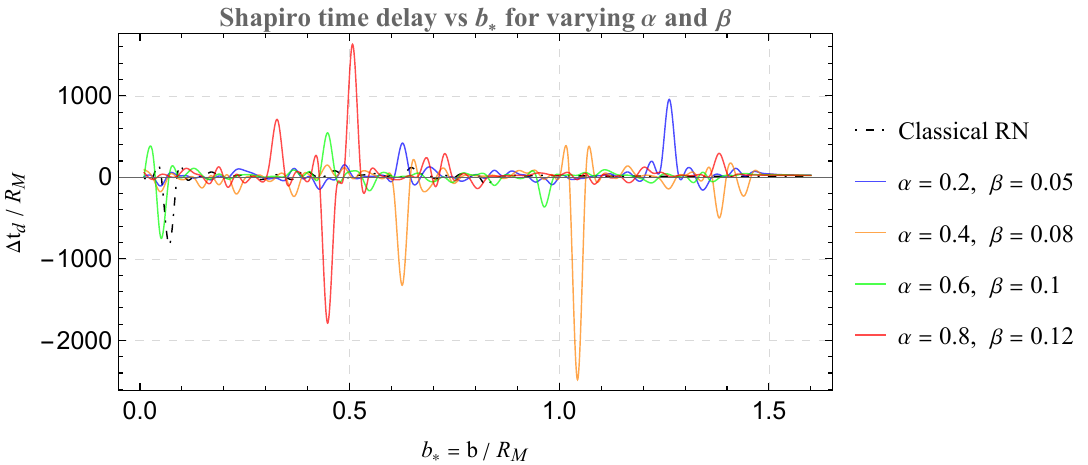}
\end{center}
\caption{\scriptsize {Shapiro time delay \(\Delta t_{\rm d}\) as a function of the normalized impact parameter \(b_* = b/R_{\rm M}\) for different values of \(\alpha\) and \(\beta\). The solid curves show the quantum-corrected results for \((\alpha, \beta) = (0.2, 0.05)\) (blue), \((0.4, 0.08)\) (orange), \((0.6, 0.10)\) (green), and \((0.8, 0.12)\) (red). The dashed black curve corresponds to the classical RN case. The delay is systematically reduced in the quantum-corrected geometry, with the reduction increasing for larger \(\alpha\) and \(\beta\). The observer is placed at \(x_A = 40\) and the source at \(x_B = 20\), with the black hole at the origin.}}
\label{fig:shapiro}
\end{figure}

\begin{figure}
\begin{center}
\includegraphics[width=7.9cm,height=4.9cm]{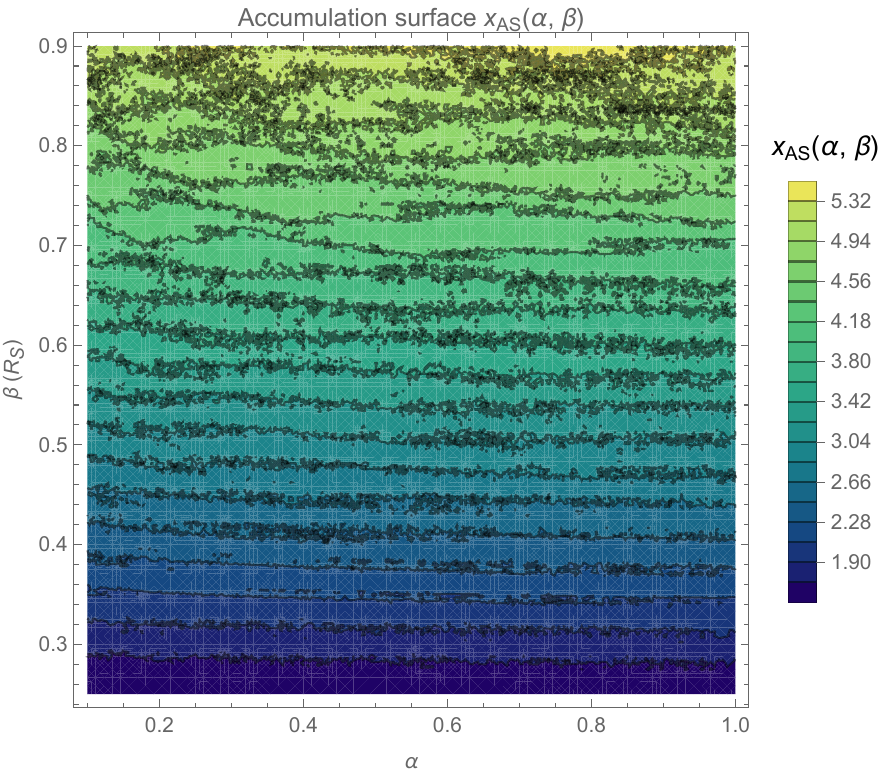}
\end{center}
\caption{ \scriptsize  The plot illustrates how the accumulation surface \(x_{\rm AS}\) depends on the parameters \(\alpha\) and \(\beta\) in our quantum-corrected black hole model. The accumulation surface marks where the effective gravitational function \(F(x, \alpha, \beta)\) vanishes. The parameter \(\alpha\) controls the strength of the quantum effect, while \(\beta\) sets the scale over which the charge distribution spreads. The color contours show how the surface shifts for different combinations of these parameters, with blue representing lower values and yellow higher values of \(x_{\rm AS}\). This visualization makes it clear how quantum corrections and the scale of the distribution influence the effective location of the accumulation surface, offering insight into the interplay between mass, charge, and quantum effects in this model.}
\label{fig-accumulation}
\end{figure}

\section{Photon Motion around Coherent Electrically Charged Quantum Black Holes}\label{sec-III}

\subsection{Inner and Outer Horizons}\label{ssSec:A}
\label{sec-hor}

In the classical RN geometry, horizon radii are identified by the condition
\begin{eqnarray}
h(R)=1+2V_{\rm RN}(R)=0,
\end{eqnarray}
where the standard solutions can be expressed as
\begin{eqnarray}
R_{\pm}^{0} = \frac{R_{\rm M}}{2}\pm\frac{1}{2}\sqrt{R_{\rm M}^{2}-4\,R_{\rm Q}^{2}}\,.
\end{eqnarray}
For \(R_{\rm M}>2R_{\rm Q}\), this yields two distinct surfaces: an outer event horizon at \(R_{+}^{0}\) and an inner (Cauchy) horizon at \(R_{-}^{0}\). At the threshold \(R_{\rm M}=2R_{\rm Q}\) the horizons coincide (the extremal case), while no horizon appears if \(R_{\rm M}<2R_{\rm Q}\).

When quantum effects are included, a new scale \(R_{\rm s}\) (the core radius) enters the geometry and modifies the effective potential. In this case, the horizon condition becomes
\begin{equation}\label{eq:horizons}
V_{\rm RN}^q(R_{\pm}^{\rm q})=-\frac{1}{2}\,.
\end{equation}
{In the limit \(R_{\rm s} \to 0^+\), we have \(R_{\pm}^{\rm q} \to R_{\pm}^{0}\), recovering the classical RN horizons.}

The solutions \(R_{\pm}^{\rm q}\) of Eq.~\eqref{eq:horizons} no longer admit a closed form in elementary functions due to the transcendental dependence introduced by the core profile. To parametrize their location efficiently, we introduce the dimensionless ratios
\begin{eqnarray}
x\equiv \frac{R}{R_{\rm M}},\quad
\alpha\equiv \frac{R_{\rm Q}}{R_{\rm M}},\quad
\beta\equiv \frac{R_{\rm s}}{R_{\rm M}}.
\end{eqnarray}
In terms of (\(x,\alpha,\beta\)), the horizon condition becomes
\begin{equation}
\tilde{h}^q(x;\alpha,\beta)=0\,,
\end{equation}
where \(\tilde{h}^q\) encodes the full quantum-corrected radial function. The roots \(x_{\pm}^{\rm q}(\alpha,\beta)\) of \(\tilde{h}^q(x)=0\) define
\begin{eqnarray}
R_{\pm}^{\rm q}=R_{\rm M}\,x_{\pm}^{\rm q}(\alpha,\beta),
\end{eqnarray}
with the ordering \(x_{-}^{\rm q}<x_{+}^{\rm q}\) in the non-extremal case.

Depending on the values of \(\alpha\) and \(\beta\), three qualitative horizon structures arise:
\begin{itemize}
    \item Two distinct horizons : (\(x_{-}^{\rm q}<x_{+}^{\rm q}\)): the geometry supports both an outer and an inner horizon, analogous to the non-extremal classical case.
    \item A single, degenerate horizon : (\(x_{-}^{\rm q}=x_{+}^{\rm q}\)): this extremal configuration occurs when parameters satisfy a critical relation such that the discriminant of \(\tilde{h}(x)\) vanishes.
    \item No horizons: \(\tilde{h}(x)\)  has no real positive roots, indicating a regular horizonless compact object.
\end{itemize}

These three regimes are illustrated in Fig.~\ref{fig:horizons_RN} as regions in the (\(\alpha,\beta\)) plane.

To map out where solutions exist, it is useful to track the asymptotic and core limits of the effective potential. Because
\begin{eqnarray}\label{eq31}
\lim_{x\to\infty}\tilde{h}(x;\alpha,\beta)=1,
\end{eqnarray}
root structure is controlled by the behavior near \(x=0\) and possible critical points. Requiring \(\tilde{h}(0;\alpha,\beta)=-\tfrac{1}{2}\) leads to a quadratic condition in \(\beta\), whose solutions can be written as
\begin{eqnarray}
\beta_{1}^{\pm}(\alpha)=\frac{1}{\sqrt{\pi}}\left[\,1\pm\sqrt{\,1-2\pi\,\alpha^{2}\,}\,\right].
\end{eqnarray}
For \(\alpha<1/\sqrt{2\pi}\), \(\beta_{1}^{\pm}\) are real and determine the transition between the three horizon patterns:
\begin{itemize}
    \item \(\beta<\beta_{1}^{-}\): two roots of \(\tilde{h}(x)=0\) exist.
    \item \(\beta_{1}^{-}<\beta<\beta_{1}^{+}\): a single root is present (extremal case).
    \item \(\beta>\beta_{1}^{+}\): no horizon forms.
\end{itemize}

These boundaries trace out an elliptical locus in the (\(\alpha,\beta\)) plane given by
\begin{equation}
2\pi\,\alpha^{2}+\Bigl(\pi\,\beta-1\Bigr)^{2}=1.
\end{equation}

For higher charge ratios \(\alpha>1/\sqrt{2\pi}\), the extremal transition disappears: only the two-horizon and horizonless regions remain, separated by a degenerate bound where \(\tilde{h}(x)\) and its derivative vanish simultaneously. 

To describe the extremal configurations more transparently, it is convenient to introduce the dimensionless parameter
\begin{eqnarray}
t \equiv \frac{x_{\rm deg}}{\beta} > 0,
\end{eqnarray}
so that the degenerate horizon can be written as \(x = \beta\,t\). With this choice, the extremality conditions take the form
\begin{eqnarray}
\tilde{h}(\beta t;\alpha,\beta)=0,\qquad
\frac{\partial \tilde{h}}{\partial x}(\beta t;\alpha,\beta)=0.
\end{eqnarray}
In this parametrization, the radial coordinate is effectively measured in units of the core radius, since \(t = R/R_{\rm s}\). This makes it easier to handle the system: instead of solving directly for the roots, one can treat \(t\) as a running parameter and express both \(\alpha\) and \(\beta\) in terms of it. In this way, one obtains a curve
\begin{eqnarray}
\bigl(\beta(t)\,,\alpha(t)\bigr),
\end{eqnarray}
which traces the boundary between configurations with two horizons and those without. This representation turns out to be particularly useful when constructing the phase diagram numerically, as shown in Fig.~\ref{fig:horizons_RN}.

From a physical point of view, it is also important to ensure that the core remains hidden inside the outer horizon. This translates into the condition
\begin{equation}
\beta < x_{+}^{\rm q}(\alpha,\beta).
\end{equation}
A simple way to probe this requirement is to evaluate the horizon equation directly at the core radius. In terms of the parametrization above, this corresponds to the special case \(x=\beta\), or equivalently \(t=1\). The condition
\begin{eqnarray}
\tilde{h}(\beta;\alpha,\beta)=0
\end{eqnarray}
then marks the transition where the outer horizon coincides with the core. Solving this relation for \(\beta\) leads to two branches,
\begin{eqnarray}
\beta_{2}^{\pm}(\alpha),
\end{eqnarray}
which separate the region where the core is safely enclosed from the one where it would lie outside the horizon. In practice, for \(\alpha \leq \alpha_{\rm crit} \simeq 0.404\), imposing \(\beta < \beta_{2}^{+}(\alpha)\) guarantees that \(R_{\rm s} < R_{+}^{\rm q}\). For larger values of \(\alpha\), whenever an outer horizon exists, it automatically surrounds the core.

Finally, whenever an inner horizon is present, it still plays the role of a Cauchy horizon. Even though quantum effects smooth out the central region and remove the classical singularity, this inner surface signals that the spacetime may remain vulnerable to the usual internal instabilities, such as mass inflation.

\subsection{Light Propagation in the Quantum-Corrected Spacetime}\label{ssSec:B}

For massless particles, the null condition \(g_{\mu\nu}\dot{x}^\mu \dot{x}^\nu = 0\) gives the radial constraint
\begin{equation}
\tilde{h}^q(x;\alpha,\beta)\,\dot{t}^2 - \frac{1}{\tilde{h}^q(x;\alpha,\beta)} \dot{x}^2 - x^2 \dot{\phi}^2 = 0,
\end{equation}
where \(\dot{x}^\mu = dx^\mu/d\tau\) is the tangent vector along the photon's worldline with \(\tau\) an affine parameter. This expression already accounts for motion in the equatorial plane, \(\theta = \pi/2\), which is sufficient due to the spacetime's spherical symmetry, allowing us to capture all the key features of the photon's trajectory.

The motion admits two conserved quantities: energy and angular momentum, defined as
\(p_t = \tilde{E}\), \(p_\phi = \tilde{L}\). It is convenient to introduce the impact parameter \(b = \tilde{L}/\tilde{E}\), which encodes how close the photon passes by the black hole. Then the evolution along the geodesic is
\begin{align}
\frac{dt}{d\tau} &= \frac{\tilde{E}}{\tilde{h}^q(x;\alpha,\beta)}, \\
\frac{d\phi}{d\tau} &= \frac{\tilde{L}}{x^2}.
\end{align}

The radial motion follows from the null condition and can be written as
\begin{equation}
\dot{x}^2 = \tilde{E}^2 - \tilde{h}^q(x;\alpha,\beta)\frac{\tilde{L}^2}{x^2} \quad \Rightarrow \quad
\frac{d\phi}{dx} = \pm \frac{1}{x \sqrt{\frac{x^2}{b^2} - \tilde{h}^q(x;\alpha,\beta)}}.
\end{equation}

The total bending angle of a photon approaching the black hole is obtained by integrating from infinity down to the closest approach \(x_1\) and back:
\begin{equation}
\Delta\phi = 2 \int_{x_1}^{\infty} \frac{dx}{x \sqrt{\frac{x^2}{b^2} - \tilde{h}^q(x;\alpha,\beta)}},
\end{equation}
where \(x_1\) is fixed by \(\dot{x}(x_1) = 0\).

A particularly important quantity is the critical impact parameter \(b_{\rm crit}\), which separates photons that are captured from those that escape. It is determined by the photon sphere radius \(X_{\rm ph}\) through
\begin{equation}
\frac{1}{b_{\rm crit}^2} - \frac{\tilde{h}^q(X_{\rm ph};\alpha,\beta)}{X_{\rm ph}^2} = 0.
\end{equation}
The photon sphere itself is found by solving
\begin{equation}
\frac{d}{dx}\left[\frac{\tilde{h}^q(x;\alpha,\beta)}{x^2}\right]\Bigg|_{x=X_{\rm ph}} = 0,
\end{equation}
which naturally generalizes the classical Schwarzschild condition to include quantum corrections.

Photon trajectories in the plane can be expressed in Cartesian coordinates as
\begin{equation}
X = x(\phi) \cos\phi, \qquad Y = x(\phi) \sin\phi,
\end{equation}
where \(x(\phi)\) satisfies
\begin{equation}
\frac{dx}{d\phi} = \pm x \sqrt{\frac{x^2}{b^2} - \tilde{h}^q(x;\alpha,\beta)}.
\end{equation}

Far from the black hole, \(\tilde{h}^q(x;\alpha,\beta) \to 1\), and both the photon sphere radius and the critical impact parameter approach their classical values. Quantum corrections slightly reduce these values, pulling the photon sphere inward and shrinking the apparent size of the shadow. This effect reflects a softening of gravity near the center, consistent with models that remove the central singularity.

The behavior of photons depends on their impact parameter \(b\) relative to \(b_{\rm crit}\):
\begin{itemize}
\item For \(b < b_{\rm crit}\), photons are inevitably captured by the black hole.
\item At \(b = b_{\rm crit}\), photons orbit near the photon sphere on an unstable circular path.
\item For \(b > b_{\rm crit}\), photons are deflected around the black hole, producing the characteristic photon ring.
\end{itemize}

This framework provides a complete description of how quantum corrections modify light propagation near coherent, nonsingular black holes and allows precise computation of shadows and lensing effects.

The gravitational field around ordinary objects such as the Earth, the Sun, or Sagittarius A$^*$ is likely influenced by the surrounding matter, so it cannot be described exactly by the quantum-corrected RN potential. Nevertheless, if a black hole of this type exists elsewhere, its presence could be revealed by carefully tracking the motion of nearby particles. As a simple illustration, we consider a particle starting at \((x_0, y_0) = (0, -1.4)\) with initial velocity \((v_x, v_y) = (0.3, 0)\), and we numerically integrate its equations of motion in the quantum-corrected potential (\ref{PotentV}).
The resulting trajectories for \(\alpha = 0.2, 0.5, 0.8\) are shown in Fig. \ref{fig-orbit}. Each value of \(\alpha\) produces a distinct path: the blue dotted line corresponds to \(\alpha = 0.2\), the black dashed line to \(\alpha = 0.5\), and the orange solid line to \(\alpha = 0.8\). The black hole is located at the origin, marked by a red dot. As \(\alpha\) increases, the trajectories become less curved, while smaller \(\alpha\) values lead to more pronounced deviations, showing how the quantum-corrected potential affects the particle's motion.

Rotation curves provide a direct way to probe deviations from standard gravitational fields. In the quantum-corrected RN potential, the circular velocity of a particle at distance \(r\) is
\begin{eqnarray}
v(r; \alpha) = \sqrt{\alpha G_N \frac{M}{r}\, \mathrm{erf}\Big(\frac{r}{R_s}\Big) - \alpha G_N \frac{Q^2}{2 r^2}\, D\Big(\frac{r}{R_s}\Big)}.
\end{eqnarray}

At small distances, the motion is dominated by the mass term, while at larger radii the contribution from the charge and the quantum correction becomes important. Lower values of \(\alpha\) lead to higher orbital speeds at large distances, showing that the quantum corrections effectively increase the gravitational pull in the outer regions. As a result, particles orbit faster than they would in the classical RN potential, highlighting the significant impact of these corrections on the motion of test bodies far from the central object.

\subsection{Shapiro Time Delay}\label{ssSec:C}

A further way to probe the influence of a quantum-corrected black hole is through the Shapiro time delay, which quantifies the excess travel time of light in a gravitational field relative to flat spacetime \cite{Gasperini:2013cru}. Within our setup, this effect follows directly from the photon geodesics and is fully controlled by the metric function \(\tilde{h}^q(x;\alpha,\beta)\).

For a photon moving in the equatorial plane, the radial component of its motion gives
\begin{equation}
\dot t = \pm \frac{\dot x}{\tilde{h}^q(x;\alpha,\beta)\sqrt{1 - \tilde{h}^q(x;\alpha,\beta) \frac{b^2}{x^2}}}\,.
\end{equation}
Consider an observer \(A\) at \(x=x_A\) sending a light signal to a second observer \(B\) at \(x=x_B\), with the black hole located between them. The photon reaches a closest approach \(x_0\) before continuing to \(B\) and then returning to \(A\). The total round-trip time is
\begin{equation}
\Delta t = 2 \left(\int_{x_0}^{x_A} + \int_{x_0}^{x_B}\right) \frac{dx}{\tilde{h}^q(x;\alpha,\beta)\sqrt{1 - \frac{x_0^2}{x^2}\frac{\tilde{h}^q(x;\alpha,\beta)}{\tilde{h}^q(x_0;\alpha,\beta)}}}\,.
\end{equation}

What is worth noting here is how the geometry enters the expression: the same function \(\tilde{h}^q\) appears both as an overall factor and inside the square root through the ratio \(\tilde{h}^q(x)/\tilde{h}^q(x_0)\). This structure keeps the integrand well-behaved at the turning point \(x_0\) while encoding the full gravitational influence along the trajectory, so the delay is not just a small correction but reflects the cumulative effect of the geometry. The Shapiro time delay, \(\Delta t_{\rm d}\), is then obtained by subtracting the flat-space travel time,
\begin{equation}
\Delta t_{\rm d} = \Delta t - 2 \left(\sqrt{x_A^2 - x_0^2} + \sqrt{x_B^2 - x_0^2}\right).
\end{equation}

We adopt Schwarzschild-type coordinates for simplicity, although other choices could be more suitable depending on the observational setup. In this setting, quantum corrections modify \(\tilde{h}^q\) in a way that softens its behavior near \(x_0\), reducing both the effective redshift and the strength of the potential barrier. As a result, the Shapiro delay is slightly suppressed compared to the classical case, consistently reflecting the smoother, nonsingular character of the underlying geometry.

{To substantiate the claim that quantum corrections reduce the Shapiro delay, we present in Fig.~\ref{fig:shapiro} the time delay \(\Delta t_{\rm d}\) as a function of the impact parameter \(b_*\) for representative values of \(\alpha\) and \(\beta\), compared with the classical RN case. The figure clearly shows that the delay is slightly suppressed in the quantum-corrected geometry, consistent with the smoother effective potential and the inward shift of the photon sphere.}

\section{Shadow Cast}\label{sec-IV}

The silhouette of a black hole encodes rich information about the near-horizon spacetime. Observations by the Event Horizon Telescope, revealing the striking shadows of M87\(^*\) and Sgr A\(^*\), have opened a new avenue for testing gravity in the strong-field regime \cite{Chen:2020aix, Chen:2022lct, Pedrotti2024, Chen:2024ibc, Wang2025a, Pantig2024, Ali2024a, Yang2025, Chen:2025jay, Al-Badawi2025, Kala2025, Chen-Yiqian2025, Pantig2025a, Wang2025b}. In particular, these images allow us to compare classical RN geometries with coherent quantum black holes, which modify the central region while preserving the overall horizon structure. Our quantum-corrected lapse function, \(\tilde{h}^q(x;\alpha,\beta)\), smoothly reduces to the familiar RN form in the appropriate limit, making it straightforward to see how quantum effects alter photon paths near the horizon. To model the emission around the black hole, we consider a thin, optically transparent accretion disk. The disk is treated as a collection of radially distributed emitters, with intensity \(j(x)\) depending only on the radial coordinate \(x\). Photons can traverse the disk freely, picking up brightness at each crossing. Repeated intersections naturally produce higher-order photon rings, layered around the central shadow. While this is a simplified model, it captures the essential physics of shadow formation and has been validated against more detailed accretion simulations \cite{Gralla:2019xty, Johnson:2019ljv}. For concreteness, we adopt a nearly face-on disk orientation, as inferred for M87\(^*\) \cite{CraigWalker:2018vam}, with the observer at \((X,Y)=(40,0)\) and the disk along \(X=0\) (Fig.~\ref{fig:trajectory}).

In many studies of black hole imaging, simple toy models are often adopted for the emission coefficient, such as those introduced in Refs.~\cite{Zeng:2020vsj, Peng:2020wun, Li:2021ypw}. Here, we follow a more physically motivated approach inspired by Ref.~\cite{Narayan:2019imo}, where the total luminosity emitted between a radius \(r\) and infinity is proportional to the gravitational energy lost by the infalling gas. We consider a spherically symmetric spacetime with metric functions expressed in terms of the dimensionless variable \(x = r / R_M\). The emission coefficient per unit solid angle, in units of (\(\mathrm{erg\;cm^{-3}\;s^{-1}\;ster^{-1}}\)) for a radiating gas at rest in the spherical spacetime, is
\begin{widetext}
\begin{eqnarray}
j(x)=
\begin{cases}
0, & x\le x_{\rm in},\\
\dfrac{1}{32\pi^2 R_M^3 x^2 \tilde{h}^q(x;\alpha,\beta)^2}
\left[
\frac{1}{x^2}\mathrm{erf}\left(\frac{x}{\beta}\right)
-\frac{2}{\sqrt{\pi}}\frac{1}{\beta x} e^{-x^2/\beta^2}
+\alpha^2\left(
-\frac{2}{x^3}\mathcal{D}\left(\frac{x}{\beta}\right)
+\frac{1}{\beta x^2}\mathcal{D}'\left(\frac{x}{\beta}\right)
\right)
\right], & x>x_{\rm in}.
\end{cases}
\end{eqnarray}

The redshift for photons observed at infinity is

\begin{eqnarray}
g(x) = \frac{1 - \dfrac{1}{x}\mathrm{erf}\left(\dfrac{x}{\beta}\right) + \dfrac{\alpha^2}{x^2}\mathcal{D}\left(\dfrac{x}{\beta}\right)}
{1 \pm \sqrt{
\left[1 - \Big(1 - \dfrac{1}{x}\mathrm{erf}\left(\dfrac{x}{\beta}\right) + \dfrac{\alpha^2}{x^2}\mathcal{D}\left(\dfrac{x}{\beta}\right)\Big)\right]
\left[1 - \dfrac{b_*^2}{x^2}\Big(1 - \dfrac{1}{x}\mathrm{erf}\left(\dfrac{x}{\beta}\right) + \dfrac{\alpha^2}{x^2}\mathcal{D}\left(\dfrac{x}{\beta}\right)\Big)\right]
}\,} 
\end{eqnarray}
\end{widetext}
where \(b_* = b / R_M\) is the dimensionless impact parameter, and the plus (minus) sign corresponds to photons moving outward (inward). This expression captures both gravitational redshift and Doppler effects of the freely falling gas.

For spherical accretion, the infinitesimal proper length along the photon trajectory is
\begin{eqnarray}
dl = R_M \, \sqrt{
1 - \frac{1}{x}\mathrm{erf}\left(\frac{x}{\beta}\right) + \frac{\alpha^2}{x^2}\mathcal{D}\left(\frac{x}{\beta}\right) + x^2 \left(\frac{d\phi}{dx}\right)^2
}  dx.
\end{eqnarray}

For a thin disk lying on the equatorial plane (\(\theta = \pi/2\)), the trajectory is parametrized by \(\phi\):
\begin{eqnarray}
dl = R_M  \sqrt{
\left[1 - \frac{1}{x}\mathrm{erf}\left(\frac{x}{\beta}\right) + \frac{\alpha^2}{x^2}\mathcal{D}\left(\frac{x}{\beta}\right)\right] \left(\frac{dx}{d\phi}\right)^2 + x^2
}  d\phi
\end{eqnarray}

The bolometric intensity observed at impact parameter \(b_*\) is
\begin{eqnarray}
I(b_*) = \int g(x)^4 \, j(x) \, dl.
\end{eqnarray}

Substituting the spherical accretion length gives
\begin{widetext}
\begin{eqnarray}
I(b_*) = \frac{1}{32\pi^2 R_M^2} &&\int
\Bigg[
\Big(1 - \dfrac{1}{x}\mathrm{erf}\left(\dfrac{x}{\beta}\right) + \dfrac{\alpha^2}{x^2}\mathcal{D}\left(\dfrac{x}{\beta}\right)\Big)^4
\frac{1}{x^2 \Big(1 - \dfrac{1}{x}\mathrm{erf}\left(\dfrac{x}{\beta}\right) + \dfrac{\alpha^2}{x^2}\mathcal{D}\left(\dfrac{x}{\beta}\right)\Big)^2}
\frac{d}{dx} \Big[1 - \frac{1}{x}\mathrm{erf}\left(\frac{x}{\beta}\right) + \frac{\alpha^2}{x^2}\mathcal{D}\left(\frac{x}{\beta}\right)\Big]
\nonumber\\&&\sqrt{
1 - \frac{1}{x}\mathrm{erf}\left(\frac{x}{\beta}\right) + \frac{\alpha^2}{x^2}\mathcal{D}\left(\frac{x}{\beta}\right) + x^2 \left(\frac{d\phi}{dx}\right)^2
}
\Bigg]  dx
\end{eqnarray}
\end{widetext}
For the thin disk, the intensity reduces to a sum over intersection points between photon trajectories and the disk:
\begin{widetext}
\begin{eqnarray}
I(b_*) = \sum_m&&
\Bigg[
\underbrace{
\Big(1 - \dfrac{1}{x_m}\mathrm{erf}\left(\dfrac{x_m}{\beta}\right) + \dfrac{\alpha^2}{x_m^2}\mathcal{D}\left(\dfrac{x_m}{\beta}\right)\Big)^4
}_{g(x_m)^4 \text{ (redshift factor)}}
\;
\underbrace{
\frac{1}{32\pi^2 R_M^3 x_m^2 \Big(1 - \dfrac{1}{x_m}\mathrm{erf}\left(\dfrac{x_m}{\beta}\right) + \dfrac{\alpha^2}{x_m^2}\mathcal{D}\left(\dfrac{x_m}{\beta}\right)\Big)}
\frac{d}{dx} \Big[ 1 - \frac{1}{x}\mathrm{erf}\left(\frac{x}{\beta}\right) + \frac{\alpha^2}{x^2}\mathcal{D}\left(\frac{x}{\beta}\right) \Big]_{x = x_m}
}_{j(x_m) \text{ (emission coefficient)}}
\nonumber\\&&\;
\underbrace{
\frac{R_M x_m^2}{b_* \sqrt{1 - \dfrac{1}{x_m}\mathrm{erf}\left(\dfrac{x_m}{\beta}\right) + \dfrac{\alpha^2}{x_m^2}\mathcal{D}\left(\dfrac{x_m}{\beta}\right)}}}_{\text{path factor from } dl}
\Bigg]
\end{eqnarray}
\end{widetext}
where \(x_m\) denotes the \(m\)-th intersection along the photon path.

Figs. \ref{fig:intensity} and \ref{fig:shadow} show how coherent electrically-charged quantum black holes influence the light observed at a distance. In Fig.~\ref{fig:intensity}, the observed intensity \(I_{\rm obs}\) is plotted against the normalized impact parameter \(b_* = b / R_M\) for {four} smearing parameters \(\beta = 0.05, 0.08, 0.10, 0.12\) and corresponding charges \(\alpha = 0.2, 0.4, 0.6, 0.8\), with the inner emission cutoff set at \(x_{\rm in} = 6\) and the outer boundary at \(x_{\rm out} = 20\). Larger \(\alpha\) values shift the intensity peaks outward and increase their magnitude, while higher \(\beta\) spreads the emission over a wider range of impact parameters, softening sharp features. Fig. \ref{fig:shadow} presents the two-dimensional intensity distribution in the observer's image plane for coherent electrically-charged quantum black holes, computed for the representative sets \((\alpha, \beta) = (0.2,0.05), (0.4,0.08), (0.6,0.10), (0.8,0.12)\). The model is evaluated in physical units with \(R_M = 2\), an inner emission cutoff at \(x_{\rm in} = 6\), and an outer boundary at \(x_{\rm out} = 20\). As seen in the figure, each panel exhibits a central dark region corresponding to the black hole shadow, surrounded by a bright ring produced by photons that orbit near the photon sphere before escaping to the observer. Increasing the charge parameter \(\alpha\) shifts this ring outward and enhances its brightness, reflecting stronger gravitational and electrostatic effects on photon trajectories. In contrast, larger values of the smearing parameter \(\beta\) lead to a more diffuse emission profile, softening the sharpness of the ring and spreading the intensity over a wider region. The overall behavior shows that quantum smearing and charge regularization modify both the size and contrast of the shadow and its surrounding emission, while preserving the global RN-like structure of the spacetime.

\section{Energy accumulation and backreaction in coherent quantum black holes}\label{sec-V}

The coherent quantum black hole geometry is determined by \(\tilde{h}^q(x;\alpha,\beta)\), with \(\alpha\) and \(\beta\) controlling the strength and radial scale of quantum modifications. To describe the build-up of energy, we model the surrounding region as a continuous distribution of thin shells. The total energy enclosed within a radius \(x\) is captured by a mass function that now explicitly depends on \(\alpha\) and \(\beta\):
\begin{equation}
\tilde{m}(x;\alpha,\beta) = \tilde{m}_0(x) + \delta\tilde{m}(x;\alpha,\beta),
\end{equation}
where \(\delta\tilde{m}(x;\alpha,\beta)\) represents the contribution from trapped radiation and quantum effects. {The perturbation \(\delta\tilde{m}(x;\alpha,\beta)\) represents contributions from both accreted matter and quantum effects, such as Hawking radiation. In the astrophysical context considered here, the dominant contribution typically arises from accreted matter, while quantum effects become important only at very late times or near the Planck scale.} The resulting spacetime metric reads
\begin{align}
ds^2 = &- \tilde{h}^q(x;\alpha,\beta)\,\sqrt{1 - \tilde{h}^q(x;\alpha,\beta)}\, \tilde{F}(x;\alpha,\beta)\, dt^2 +\frac{R_M^2}{\tilde{F}(x;\alpha,\beta)}\, dx^2 \nonumber \\
  &+ R_M^2 x^2 \left(d\theta^2 + \sin^2\theta, d\phi^2 \right),
  \end{align}
  with
  \begin{equation}
  \tilde{F}(x;\alpha,\beta) = 1 - \frac{2\tilde{m}(x;\alpha,\beta)}{R_M x}.
  \end{equation}
  The surface where energy accumulates is naturally defined by \(\tilde{F}(x_{\rm AS};\alpha,\beta) = 0\), which acts as a dynamical, energy-sensitive boundary.

To see how this surface responds to additional energy, we perturb the mass function,
\begin{equation}
\tilde{m}(x;\alpha,\beta) = \tilde{m}_0(x;\alpha,\beta) + \delta \tilde{m}(x;\alpha,\beta),
\end{equation}
and allow the accumulation surface to move accordingly,
\begin{equation}
x_{\rm AS}(\alpha,\beta) = x_{\rm AS}^0(\alpha,\beta) + \delta x_{\rm AS}(\alpha,\beta).
\end{equation}
A first-order expansion of the condition \(\tilde{F}(x_{\rm AS};\alpha,\beta)=0\) gives
\begin{equation}
\delta x_{\rm AS}(\alpha,\beta) \approx
2 \, \frac{\delta \tilde{m}(x_{\rm AS};\alpha,\beta)}{R_M} \, \Big( \frac{\partial \tilde{F}}{\partial x}\Big)^{-1}_{x_{\rm AS}}.
\end{equation}
Since \(\partial_x \tilde{F}(x_{\rm AS};\alpha,\beta)\) is positive at the outermost root, any increase in energy pushes the accumulation surface outward. The parameters \(\alpha\) and \(\beta\) directly shape this response: a larger \(\alpha\) strengthens the deformation, increasing the sensitivity of the surface to incoming energy, while a larger \(\beta\) spreads the effect over a wider region, producing a smoother, more gradual outward shift. The function \(\tilde{h}^q(x;\alpha,\beta)\) also plays a crucial role. It keeps the redshift finite near the core, preventing divergences typical of classical solutions. This regularization is more pronounced for larger \(\beta\), which smooths out sharp gradients, while larger \(\alpha\) enhances the trapping of energy without causing singular behavior.

Unlike horizonless objects where energy trapped near light rings often causes the trapping region to shrink, here the effective horizon expands as energy accumulates. This expansion highlights a key difference in backreaction: the interplay between \(\tilde{h}^q(x;\alpha,\beta)\) and \(\tilde{m}(x;\alpha,\beta)\) governs both the trapping efficiency and the dynamical response. In other words, \(\alpha\) sets the strength of the effect, while \(\beta\) determines how it is distributed across the spacetime. The analysis presented is perturbative, and while it demonstrates that the accumulation surface remains well-behaved under moderate energy influx, a full exploration of the nonlinear regime---particularly for extreme \(\alpha\) and \(\beta\)---remains an open avenue for future investigation. The contour plot of \(x_{\rm AS}(\alpha, \beta)\) (Fig. \ref{fig-accumulation}) reveals how the location of the accumulation surface shifts as a function of the quantum correction strength and the characteristic scale of the charge distribution. We see that increasing \(\alpha\), which amplifies the quantum contribution to the mass, generally pushes the accumulation surface outward, while larger values of \(\beta\) spread the effect of the charge over a wider region, slightly reducing the surface radius. This behavior highlights the delicate balance between classical gravitational attraction, the charge distribution, and quantum effects. Overall, the figure demonstrates that even modest quantum corrections can noticeably modify the effective structure near the black hole, providing insight into how such corrections influence horizon-like surfaces and potentially observable properties.

\section{Discussion and conclusions}\label{sec-VI}

In this work, we revisited the RN geometry from the perspective of coherent quantum black holes, with the aim of building a description that remains well-behaved at short distances while preserving the correct large-scale physics. The key step was to replace the sharp ultraviolet cutoff used in earlier constructions with a smooth Gaussian suppression. This seemingly modest change turns out to be essential: it removes the spurious oscillations in the effective potential and leads to a geometry that is smooth, stable, and physically interpretable. At the same time, the classical RN solution is recovered in the appropriate limit, so the model remains anchored to known results.

Looking more closely at the central region, the structure that emerges is not one of a singularity simply disappearing, but rather one where its nature is softened. The metric becomes analytic at the origin, and depending on how the parameters are related, the spacetime can either be {Ricci-regular} or exhibit a milder, integrable singularity. In the latter case, curvature scalars still diverge, but in a controlled way that allows physical quantities such as total energy to remain finite. {We have clarified that for generic parameters, even at the critical value \(R_{\rm s}^*\), the Kretschmann scalar retains a \(1/r\) divergence, so the center is not fully regular in the sense of finite curvature invariants. This suggests that quantum effects do not necessarily erase singular behavior altogether, but instead reshape it into something less pathological.} This behavior is reflected in the effective matter content associated with the geometry. The spacetime can be interpreted as being sourced by an anisotropic fluid whose properties are directly tied to the regularization procedure. Near the core, the energy density and radial pressure grow, while the tangential pressure stays finite, pointing to a strongly anisotropic configuration. Even so, the overall energy distribution remains well-defined when integrated over space. The Misner-Sharp mass further clarifies this behavior by showing how the mass builds up smoothly from the center and approaches the expected asymptotic value at large distances.

On a global level, the introduction of the quantum scale \(R_{\rm s}\) enriches the structure of the spacetime. Instead of a single fixed horizon pattern, one finds a range of possibilities depending on the parameters: configurations with two horizons, extremal cases with a single degenerate horizon, and even horizonless geometries. The transition between these regimes can be clearly mapped in terms of the dimensionless parameters \(\alpha\) and \(\beta\). This makes it evident that quantum corrections do more than modify the core---they also influence the overall causal structure. The impact of these modifications becomes particularly clear when looking at photon motion. The quantum-corrected potential slightly weakens the gravitational pull near the center, which in turn shifts the photon sphere inward and reduces the critical impact parameter. As a result, light bending and time delay effects are subtly altered. These changes are small far from the object but become noticeable closer in, where the quantum corrections are most relevant.

These features carry over naturally to observational signatures. The black hole shadow still has the familiar appearance---a dark central region surrounded by a bright ring---but its detailed structure is modified. Increasing the charge tends to push the emission outward and enhance the brightness, while increasing the smearing scale spreads the emission more evenly and softens sharp features. The overall effect is a shadow that is slightly smaller and less sharply defined than in the classical case, though still broadly consistent with current observations. We also examined how the system responds to additional energy. By allowing the mass function to vary, we identified an accumulation surface that behaves like an effective, dynamical boundary. When energy is added, this surface moves outward, indicating that the region capable of trapping radiation expands. This behavior contrasts with some alternative models and highlights the role played by the quantum-corrected geometry. The parameters \(\alpha\) and \(\beta\) again control how strong and how smooth this response is, and importantly, the entire process remains free of divergences within the regime considered.

Taken together, these results show that it is possible to construct a quantum-corrected, electrically charged black hole model that is both mathematically controlled and physically transparent. The same framework consistently connects the behavior at the core, the global structure, the motion of light, and the observable appearance of the object. There are still open questions---particularly regarding stability and fully nonlinear dynamics---but the picture that emerges is a coherent one, where quantum effects modify the classical geometry in a smooth and meaningful way rather than introducing new pathologies.

\section*{Acknowledgments}
{We thank the anonymous reviewers for their valuable comments and detailed suggestions, which have greatly helped to improve the clarity and quality of this work. }

\section*{Conflict Of Interest statement } 
No conflict of interest declared by the authors.

\section*{Data Availability Statement}  
No data were created or analyzed in this study.

\bibliography{references}

\end{document}